\documentclass[twocolumn,showpacs,preprintnumbers,amsmath,amssymb]{revtex4-1}
\usepackage{graphicx}
\usepackage{dcolumn}
\usepackage{amsfonts,amsmath,amssymb,bm}

\begin{document}

\title{\bf {Comment on \textquotedblleft Dynamics distortions and polaronic effects in the paramagnetic state of 
$La_{0.8}Ba_{0.2}Mn_{1-x}Al_xO_3$\textquotedblright[J. Appl. Phys. 2014, 115, 223905]}}
\author{J.~Ardashti~Saleh}
\author{I.~Abdolhosseini Sarsari$^{*}$}
\author{P.~Kameli}
\author{H.~Salamati}
\affiliation{Department of Physics, Isfahan University of 
Technology, Isfahan, 84156-83111, Iran/
$^{*}$abdolhosseini@cc.iut.ac.ir}


\maketitle

\noindent 
Recently, the authors of ref.~\cite{Narreto2014}, reported dynamics distortion
and polaronic effects in the paramagnetic state of $La_{0.8}Ba_{0.2}Mn_{1-x}Al_xO_3$
manganites. In this comment we want to complete  the phase diagram(Fig3\cite{Narreto2014}) 
and point out of  occurring spin-glass state in this manganite 
which is not intended for x=0.15 doping.

The authors of ref.~\cite{Narreto2014} reported a decrease of 
antiferromagnetism at higher Al doping (x\textgreater0.15) at low temperatures. 
They proposed transition to a spin glass (SG) state due to higher spin disorder.
Also they claimed ferromagnetic metallic
behavior at x=0.15 doping, while this sample shows 
a very weak metal-insulator transition in Fig2a~\cite{Narreto2014}, and 
at lower temperatures, is completely insulator. 
Thus, we examined if it is in spin glass 
(SG) state due to its insulation behavior at low temperatures. 
Also this state at lower temperature in x=0.15 
sample, has not been 
considered in to the phase diagram (Fig3\cite{Narreto2014}).

In other words, their phase diagram is not
complete. Therefore we have investigated the subject of SG in this comment.
Bulk sample of x=0.15 in our compound, were synthesized using sol-gel method.

One of the features of SG system is the dependence of
its ac susceptibility on the applied field 
and frequency. 
In this sample, this  behavior results
in a sharp drop in the real part of the ac
susceptibility at low temperatures, and the appearance
of a peak in its imaginary part.
The frequency-independent peaks named as Hopkinson peaks are typical
feature in many ferromagnetic materials, but the 
second peak positions, shift to higher temperatures with
increasing frequency. So, to verify SG state presence,
we have  measured the imaginary ac susceptibility of the 
$La_{0.80}Ba_{0.20}Mn_{0.85}Al_{0.15}O_3$ (x=0.15) sample on the constant field and different frequencies.

Figure 1 shows the temperature dependence of the 
imaginary part (Out of phase) of the ac susceptibility at different frequencies.
The temperature dependence of ac susceptibility of this sample measured in an applied field of
10 Oe, after cooling in the absence of zero field (ZFC). As can be seen from Figure 1,
there is SG behavior below a certain temperature (see inset in Fig.1).

The first peak temperature that is frequency independent, represents the Curie temperature, while
the second peak temperature, represent the freezing temperature $T_f$, that shifts towards higher temperatures
with increasing frequency. This change by frequency is characteristic of SG phase. For further
investigation of the SG nature at x=0.15 doping, we have checked the $T_f$'s dependence on frequency 
by conventional critical slowing down model which is as follows\cite{Liu2014}:
$f=f_0(\frac{T_f-T_g}{T_g})^{z\nu}$

\begin{figure}[htbp]
\includegraphics*[width=0.30\textwidth]{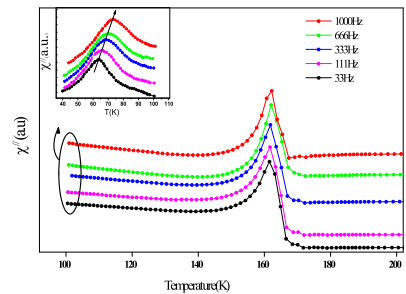}  
\caption{\label{Scheme} The inset shows the evolution of the peak by the increase of the frequency.} 
\end{figure}

where, $T_g$ is the dc value of $T_f$ for $f\rightarrow0$, $f_0$ is a constant in order of $10^9-10^{13}$and $z\nu$ is dynamic critical exponent. 
The Fig. 2 shows the best fit of this model. The estimated values are within the 
realm of three dimensional SGs that dose not coincide with ref.~\cite{Narreto2014}

\begin{figure}[htbp]
\includegraphics*[width=0.30\textwidth]{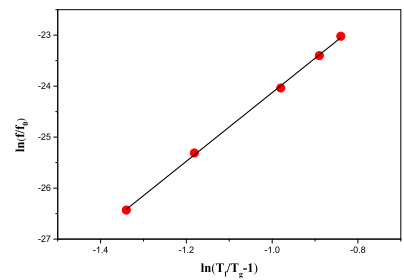}  
\caption{\label{Scheme} Ln-Ln plot of the reduced temperature $(T_f/T_g-1)$ versus 
frequency for x=0.15 sample.} 
\end{figure}
\bibliography{Comment.bib}
\end{document}